\documentclass[11pt,a4paper]{article}
\pdfoutput=1
\usepackage{jheppub}

\makeatletter
\def\@fpheader{\relax}
\makeatother

\usepackage{fancyhdr}
\usepackage{amsmath,amsthm,amssymb}
\usepackage{enumerate}
\usepackage{color}
\usepackage{graphicx}
\usepackage{caption}

\def\x{\mathrm{x}}
\def\p{\mathrm{p}}

\def\beq{\begin{equation}}
\def\eeq{\end{equation}}
\def\bea{\begin{eqnarray}}
\def\eea{\end{eqnarray}}

\def\bwt{\begin{widetext}}
\def\ewt{\end{widetext}}

\begin{document}

\title{Casimir Effect for Nonlocal Field Theories with Continuum Massive Modes}

\author{Mehdi Saravani}
\affiliation{School of Mathematical Sciences, University of Nottingham, University Park, Nottingham, UK}
\emailAdd{mehdi.saravani@nottingham.ac.uk}

\abstract{
In this paper, we study the Casimir force for a class of Lorentzian nonlocal field theories \cite{Saravani:2015rva, Belenchia:2014fda}. These theories include a continuum of massive excitations. In this regard, the effect of continuum massive modes on Casimir force is of interest. We focus on the simplest case of two absorbing parallel planes in 1+1 dimensions, and we show that, unlike local field theories, the thickness of the absorbing ``walls'' changes the value of Casimir force.  
}

\date{}
\maketitle

\section{Introduction}

Nonlocal quantum field theories (NLQFT) have been introduced in the past for various reasons, mainly to cure divergences of quantum field theories \cite{PhysRev.79.145, PhysRevD.41.1177}. Recently, a class of NLQFTs has been revisited as the low energy limit of quantum field theories on causal sets. Causal set is an approach to quantum gravity that replaces the spacetime continuum with a causet, a discrete structure which keeps the notion of causality \cite{PhysRevLett.59.521}. 
In these NLQFTs, the d'Alembertian operator $\Box=\eta^{\mu\nu}\partial_\mu\partial_\nu$ is replaced by a nonlocal operator $f(\Box)$ where $f$ is a non-analytic function. The modification is designed such that one recovers a local field theory in the low energy limit \cite{Sorkin:2007qi}. So, we can think of these theories as high energy modifications of local field theories. 
If Causal set describes the fundamental structure of spacetime, we expect this class of NLQFTs to be a reasonable approximation in the low energy (sub-planckian) regimes. As a result, studying these theories and their predictions could be an important step towards understanding fundamental structure of spacetime. 

The nonlocality introduced here is a Lorentzian nonlocality, i.e. nonlocality in spacetime, and it should be differentiated from Lorentz breaking NLQFTs. As a result, this work is a complimentary study of nonlocality in comparison to the phenomenological studies of Lonretz breaking NLQFTs \cite{PhysRevLett.116.061301}. 

An important feature of these theories is that their spectrum includes a continuum of massive excitations. This property has been studied thoroughly in \cite{Saravani:2018rwm}, and continuum massive modes are proposed to constitute cosmological dark matter in \cite{Saravani:2015rva, Saravani:2016enc}. Leading order modifications to known physical processes due to nonlocality and continuum modes have been studied in \cite{Belenchia:2016sym}. In this paper, we focus on the modification to Casimir force between two parallel planes.
  
In order to calculate Casimir force for parallel planes in local quantum fields, there are different physical setups that leads to the same final result. Here, we mention two. In one case, the two (infinitely) thin parallel planes are positioned within a distance. In this case, the boundary conditions are imposed such that the field vanishes on the planes, but it can take any value anywhere else. In another setup, we put the field in a box and we assume that the value of the field outside the box is zero. Although these two setups are physically different, the value of Casimir force with the same separation in both cases is equal. The reason for that stems from the locality of the field.

It is not clear that these setups for a nonlocal field theory amount to the same result, since a nonlocal field theory may, in general, cares about the thickness of the walls. In fact, we show that they end up giving different values for Casimir force. The calculation with the first setup (with infinitely thin walls) has been done in \cite{Saravani:2018rwm}. Here, we find the answer for the second setup (infinitely thick walls). 
Before doing that we may ask which one of the setups is relevant for experiments? Given that we expect the nonlocality length scale to be the smallest length scale in experiments, it seems that the second setup where the thickness of the walls are much larger than nonlocality length is the one better reflecting a physical experiment. 

We use the following notation: $p\cdot q = \eta_{\mu\nu}p^\mu q^\nu$ where $\eta_{\mu\nu}$ is the Minkowski metric with mostly positive sign.  In this paper, we perform the calculations in $D=1+1$ dimensions and $x^\mu = (t, \mathrm{x})$, $p^\mu = (p^0, \mathrm{p})$.

\subsection{Zero point energy}

The zero point energy of a quantum field in the vacuum is infinite. In order to derive the Casimir force between two (absorbing) plates in the vacuum of a free nonlocal scalar field theory, we need to calculate how much this infinite energy changes when plates are put in place, with boundary conditions $\phi(\mathrm{x}<0)=\phi(\mathrm{x}>a)=0$ where $a$ is the separation between planes.

In order to do this, we will follow the idea presented in \cite{AMBJORN19831}. The overall idea is to calculate the zero point energy through the partition function of the theory. Here, we present a summary of the calculations. 

Given the time-ordered two point correlation function of a free theory (with no boundary), we can perform the following Fourier decomposition
\beq
\langle 0| T\phi(x) \phi(y)|0\rangle= -i  \int \frac{d^Dp}{(2\pi)^D}G_F(p)e^{ip\cdot (x-y)}
\eeq 
where $G_F(p)$ is the time-ordered two point function in the momentum space. Imposing the boundary conditions only allows certain wave numbers $\p_n=\frac{n\pi}{a}$ and replaces the Fourier transform with a Fourier series. 

The partition function ($Z$) of the theory is given through making time imaginary (with period $\beta$), or equivalently by sending $p^0\rightarrow ip^0$.
Using $G_E(p)$ as the wick rotated Feynman propagator, i.e. $G_E(p^0,\p)=G_F(ip^0,\p)$, we get at low temperature limits ($\beta \rightarrow \infty$)
\beq
\ln Z=-\frac{1}{2}\frac{\beta}{2\pi}\int dk\sum_{n=0}^{\infty}\ln G_E\left(k,\frac{n\pi}{a}\right).
\eeq
As a result, the zero point energy is given by 
\beq\label{E0}
E=-\frac{1}{4\pi}\int_{-\infty}^\infty dk\sum_{n=0}^{\infty}\ln G_E\left(k,\frac{n\pi}{a}\right).
\eeq
Thus, we only need to know the Feynman correlation function of the theory.

\subsection{NLQFT correlation functions} 
For the calculation of zero point energy, as described above, we only need the Feynman propagator. For a full list of correlation functions and more discussion on the nonlocal field theory, we refer the readers to \cite{Saravani:2018rwm}. 

The Feynman propagator is given by
\beq\label{G_F}
G_F(p)=\int dm^2 \rho(m^2) \frac{1}{p^2+m^2-i \epsilon}
\eeq
where $p^2=\eta_{\mu\nu}p^\mu p^\nu$. This is very similar to the two point correlation function of an interacting local field theory. However, this similarity is just in the form of the two point function. In particular, the nonlocal theory considered here is free, i.e. all higher order correlation functions are fixed by the two point function. 

The spectral function, $\rho(m^2)$, has the following form
\beq
\rho(m^2) = \delta(m^2) + \tilde \rho(m^2)
\eeq
where $\tilde \rho(m^2)$ is a smooth function, free from singularities. The information about nonlocality is encoded in $\tilde \rho$ function. In particular, in the local limit this function vanishes and we recover a free massless scalar field theory. In fact, this property shows this nonlocal field theory is a high energy modification of a massless scalar field theory. Thus, we expect the zero order contribution to the Casimir force to be of a massless scalar field. 

In order to perform the Casimir force calculations, we need to choose an explicit form for $\tilde \rho$ function. In this paper, we consider two analytic functions. They will make calculations easier to handle, while they satisfy the general properties we expect this function to satisfy. They are,
\bea
\tilde \rho_1(m^2) &=& l^2 e^{-m^2 l^2}\\
\tilde \rho_2(m^2) &=& l^2 \theta(1-ml)
\eea
where $l$ represents the nonlocality length scale and $\theta$ is the Heaviside function. We show that the leading order modification to the Casimir force (in the limit $l/a \ll 1$) is the same for both cases.

\section{Riemann Zeta regulator}

In this section, we find the leading order modification to the Casimir force using Riemann Zeta function regulator.

If we define
\beq\label{E_analytic}
E(s)=-\frac{1}{4\pi}\int_{-\infty}^\infty dk\sum_{n=1}^{\infty}\ln G_E\left(k,\frac{n^{-s}\pi}{a}\right),
\eeq
the Casimir energy is given by $E=E(-1)$\footnote{Note that $n=0$ term in eq. \eqref{E0} does not depend on $a$ and thus does not contribute to the Casimir force.}, where we evaluate the right hand side by analytic continuation.

Let's start with the case $\tilde \rho_1(m^2) = l^2 e^{-m^2 l^2}$.
Replacing this function into eq. \eqref{G_F} and performing the Wick rotation, we get
\beq
G_E(p)=\frac{1}{p_E^2}-l^2e^{-l^2 p_E^2}\mbox{Ei}[-l^2p_E^2]
\eeq
where $p_E^2=\delta_{\mu\nu}p^\mu p^\nu$ and Ei is the Exponential integral function. In order to perform the integral in eq. \eqref{E0}, first we expand $\ln G_E(p)$ in powers of $l$
\beq\label{expansion}
\ln G_E(p)=-\ln p_E^2-l^2p_E^2\left[\gamma_E+\log(l^2 p_E^2)\right]+\cdots
\eeq
where $\gamma_E$ is the Euler gamma constant. From now on, we only keep the leading order contribution in $l$ and drop $\cdots$ terms. 
Substituting above in eq. \eqref{E_analytic}, we get
\beq
E(s)=\frac{1}{4\pi}\sum_{n=1}^{\infty}\int_{-K}^K dk \ln \left[k^2+\left(\frac{n^{-s}\pi}{a}\right)^2\right]+l^2\left(k^2+\left(\frac{n^{-s}\pi}{a}\right)^2\right)\left(\gamma_E+\ln\left[l^2 k^2+l^2\left(\frac{n^{-s}\pi}{a}\right)^2\right]\right)
\eeq 
where $K$ is a cut-off taken to infinity at the end of the calculation. Performing the above integral and keeping only terms surviving in $K\rightarrow \infty$ which depends on $a$, we get
\beq
E(s)=\frac{1}{4\pi}\sum_{n=1}^\infty n^{-2s} (\pi l/a)^2\left[-2+2\gamma_E+4\ln(K l)\right]K +2\pi^2 n^{-s}/a+\frac{4}{3}l^2\pi^4n^{-3s}/a^3.
\eeq
Now, we use Riemann zeta function to evaluate the sum above. This yields,
\beq\label{casimir_energy_s}
E(s) =\frac{1}{4\pi}\left\{\zeta(2s)(\pi l/a)^2\left[-2+2\gamma_E+4\ln(K l)\right]K +2\pi^2 \zeta(s)/a+\frac{4}{3}l^2\pi^4\zeta(3s)/a^3\right\}.
\eeq
At $s=-1$,
\beq\label{casimir_energy}
E=-\frac{\pi}{24a}+\frac{\pi^3l^2}{360a^3}.
\eeq
Note that $\zeta(-2)=0$, $\zeta(-1)=-1/12$ and $\zeta(-3)=1/120$. The first term above is the value of the Casimir energy of a massless scalar field, and what we expect to be the leading order contribution. The second term is the modification due to nonlocality effects.

Similar steps for $\tilde \rho_2(m^2) = l^2 \theta(1-ml)$, corresponding to
\beq
G_E(p)=1/p_E^2+l^2\int_0^{1/l^2} \frac{dm^2}{p_E^2+m^2}=1/p_E^2+l^2\ln\left(\frac{l^2p_E^2+1}{l^2p_E^2}\right),
\eeq
would result in the same value for the Casimir energy, eq. \eqref{casimir_energy}.

\section{Mass discretization}

In the previous section, we have presented a rather straightforward manner to calculate the Casimir energy for a nonlocal field theory. 
In this section, we follow a different method to calculate the Casimir energy. This method is more involved but it does not rely on Riemann Zeta function regulators, and it provides an independent calculation to validate eq. \eqref{casimir_energy} result. For simplicity, we consider the second case above, $\tilde \rho_2(m^2) = l^2 \theta(1-ml)$.
The Wick rotated Green's function is given by
\beq
G_E(p)=\frac{1}{p_E^2} +\int_0^{l^{-2}} dm^2 \frac{l^2}{p_E^2 +m^2}  
\eeq
If we discretize the above integral over mass, it yields
\beq\label{Gdisc}
G_E^{(N)}(p)=\frac{1}{p_E^2}+\frac{1}{N}\sum_{k=1}^{N}\frac{1}{p_E^2+k \Delta}
\eeq
where $l^{-2} =N\Delta$ and $\Delta$ is the discretization step in $m^2$. One can verify directly that $\lim_{N\rightarrow \infty}G_E^{(N)}=G_E$. 

First, let us show that $G_E^{(N)}$ can be written in the following manner
\beq\label{Gdisc2}
G_E^{(N)}=A\frac{(p_E^2 +u_1 \Delta)(p_E^2 +u_2 \Delta)\cdots (p_E^2 +u_N \Delta)}{p_E^2(p_E^2 +\Delta)(p_E^2 +2 \Delta)\cdots (p_E^2 +N \Delta)}
\eeq
where A is a constant (independent of $p$) and $u_k$'s are zeros of the following function 
\beq\label{fdef}
h(u)=\frac{1}{u}+\frac{1}{N}\sum_{k=1}^{N}\frac{1}{u-k}
\eeq
and $k-1<u_k<k$. 

This can be shown by defining $u=-\frac{p_E^2}{\Delta}$ and noting that $h(u)=-\Delta G_E^{(N)}(p)$. Taking the common denominator in eq. \eqref{fdef}, we get
\beq
h(u)=\frac{P(u)}{u(u-1)(u-2)\cdots (u-N)}
\eeq
where $P(u)$ is a polynomial of degree $N$, thus it has at most $N$ real zeros. On the other hand, one can easily verify from eq. \eqref{fdef}, 
\beq
\lim_{u\rightarrow k-1^+}h(u)=+\infty,~~~\lim_{u\rightarrow k^-}h(u)=-\infty,~~ k \in \{1,2,\cdots, N\}.
\eeq
This shows that $h$ (and $P$) has at least one zero ($u_k$) between $k-1$ and $k$ for $k \in \{1,2,\cdots, N\}$. Combining this with the fact that $P(u)$ has at most $N$ real zeros concludes that $P(u)$ has exactly $N$ zeros $u_k$, such that $k-1<u_k<k, ~ k \in \{1,2,\cdots, N\}$. This means that 
\beq
h(u)=A' \frac{(u-u_1)(u-u_2)\cdots(u-u_N)}{u(u-1)(u-2)\cdots(u-N)}
\eeq
which results into eq. \eqref{Gdisc2} by replacing $u=-\frac{p_E^2}{\Delta}$. 

Substituting eq. \eqref{Gdisc2} in eq. \eqref{E0} and expanding the $\ln$ function, we notice that each term corresponds to the Casimir energy of a massive local field\footnote{For a local field with mass $m$, $G_E(p)=\frac{1}{p_E^2+m^2}$.}. This gives
\beq\label{casimir_disc}
E=E^C_0+\lim_{N\rightarrow \infty}\sum_{k=1}^N E^C_{k \Delta}-E^C_{u_k \Delta}
\eeq
where $E^C_{m^2}$ is the Casimir energy associated to a local scalar filed with mass $m$, given by (in $d+1$ dimensional spacetime)
\beq
E^C_{m^2}=\frac{S_{d-1}}{2(2\pi)^d}\int_0^\infty dk~ k^{d-1}\ln\left[1-e^{-2\sqrt{m^2+k^2}a}\right],
\eeq
where $S_{d-1}$ is the area of a unit $d-1$ sphere. Also note that 
\beq
u_k<k\rightarrow E^C_{k\Delta}>E^C_{u_k \Delta},
\eeq
meaning that the Casimir energy in the nonlocal theory is higher than the Casimir energy of a massless field, $E>E_0^C$, so far consistent with the result of the previous section.

\subsection{Analytical evaluation}

In principle, one can evaluate eq. \eqref{casimir_disc} numerically and checks whether the result is consistent with eq. \eqref{casimir_energy}. We follow that in the next section.

In this section, we derive the leading order term to eq. \eqref{casimir_disc} for small values of $l/a$ analytically. We make use of the fact that for large masses ($m\gg a^{-1}$), the Casimir energy $E^C_{m^2}$ is exponentially suppressed. As we will show, this means that the leading contribution to eq. \eqref{casimir_disc} comes from $m \lesssim a^{-1}$.
 
Since $k-1<u_k<k$ and $E^C_{m^2}$ is a monotonically increasing function in terms of $m$,
\beq
E^C_{(k-1)\Delta}<E^C_{u_k \Delta}<E^C_{k \Delta}.
\eeq
As a result,
\beq
0<\sum_{k=N_0+1}^{N}E^C_{k\Delta}-E^C_{u_k\Delta}<\sum_{k=N_0+1}^{N}E^C_{k\Delta}-E^C_{(k-1)\Delta}=E^C_{N\Delta}-E^C_{N_0\Delta}
\eeq
for any value of $N_0 < N$. If we choose
\beq\label{N_0}
N_0=\alpha N l/a
\eeq
where $\alpha=\mathcal{O}(1)$ number to make the right hand side an integer, we get
\beq
0<\sum_{k=N_0+1}^{N}E^C_{k\Delta}-E^C_{u_k\Delta}<E^C_{l^{-2}}-E^C_{\alpha/al}.
\eeq
Using the above inequality, we get
\beq\label{casimir_inequality}
\lim_{N\rightarrow \infty}\sum_{k=1}^{N_0} E^C_{k \Delta}-E^C_{u_k \Delta}<E-E^C_0<E^C_{l^{-2}}-E^C_{\alpha/al}+\lim_{N\rightarrow \infty}\sum_{k=1}^{N_0} E^C_{k \Delta}-E^C_{u_k \Delta}
\eeq
Now, we evaluate 
\beq\label{Idef}
I=\lim_{N\rightarrow \infty}\sum_{k=1}^{N_0} E^C_{k \Delta}-E^C_{u_k \Delta}.
\eeq
Note that the maximum squared mass in this sum is given by $N_0 \Delta = \frac{\alpha}{la}$ which is independent of $N$.

Let us define $-1<\delta u_k<0$ as follows
\beq
\delta u_k= u_k-k.
\eeq
For large values of $N$ (small $\Delta$), we can use the following approximation
\beq\label{delta_casimir}
E^C_{k\Delta} - E^C_{u_k \Delta} \approx -\frac{dE^C}{dm^2}\bigg\rvert_{k\Delta}\delta u_k\Delta.
\eeq

\begin{figure}
\includegraphics[width=\textwidth]{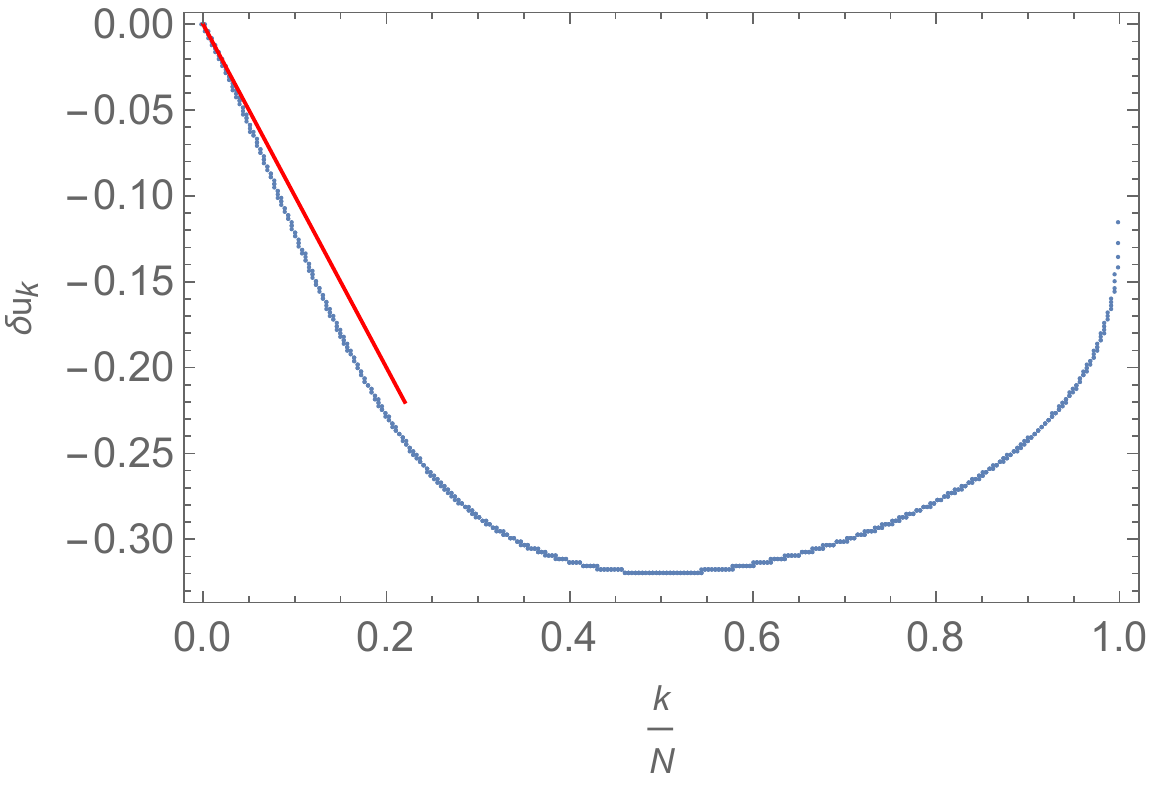}
\caption{The blue curve, consisting of $N=1000$ points, shows $\delta u_k = u_k - k$ versus $k/N$. The red line shows $-k/N$.}\label{fig:h}
\end{figure}

Fig. \ref{fig:h} shows $\delta u_k$ for $N=1,000$. In order to calculate $I$ for small values of $l/a$, only the values of $\delta u_k$ for small $k/N$ is relevant. This comes from the fact that the maximum value of $k$ appearing in the sum for $I$ (eq. \eqref{Idef}) is $N_0$ and $\frac{N_0}{N} = \frac{\alpha l}{a}$. This is, in fact, the reason behind introducing $N_0$ as in eq. \eqref{N_0}. 

From this figure, we can see that for small values of $k/N$, $\delta u_k \approx -k/N$. Let us prove this. 

As $u_k=k+\delta u_k$ is a zero of the function $h(u)$, we have
\beq\label{deltauk}
\frac{1}{k+\delta u_k}+\frac{1}{N}\left(\frac{1}{\delta u_k}+\frac{1}{\delta u_k+1}+\sum_{n=1}^{k-2} \frac{1}{k-n+\delta u_k}+\sum_{n=k+1}^{N} \frac{1}{k-n+\delta u_k}\right)=0.
\eeq
Furthermore, 
\bea
&&\lim_{N\rightarrow \infty}\frac{1}{N}\sum_{n=1}^{k-2} \frac{1}{k-n+\delta u_k}=0,\\
&&\lim_{N\rightarrow \infty}\frac{1}{N}\sum_{n=k+1}^{N} \frac{1}{k-n+\delta u_k}=0.
\eea 
Using the above results in eq. \eqref{deltauk} for a {\it fixed} value of $k$, the only solution consistent with $-1<\delta u_k<0$ for large values of $N$ is 
\beq
\delta u_k\approx -k/N.
\eeq

Substituting this in eq. \eqref{delta_casimir} yields
\beq
E^C_{k\Delta} - E^C_{u_k \Delta} \approx \frac{dE^C}{dm^2}\bigg\rvert_{k\Delta}\frac{k}{N}\Delta.
\eeq
Finally, 
\bea
I&=&\lim_{N\rightarrow \infty}\sum_{k=1}^{N_0} E^C_{k \Delta}-E^C_{u_k \Delta}\notag\\
&=&\lim_{N\rightarrow \infty}\sum_{k=1}^{N_0} \frac{dE^C}{dm^2}\bigg\rvert_{k\Delta}\frac{k}{N}\Delta\notag\\
&=& \int_0^{\alpha/la}dm^2 \frac{dE^C}{dm^2} m^2 l^2 \notag\\
&=&(m^2 l^2 E^C_{m^2})\bigg\rvert^{\alpha/la}_{0}-l^2 \int_0^{\alpha/la}dm^2 E^C_{m^2}.
\eea
Evaluating above and keeping the leading order terms in $l$, we get
\beq
I = -l^2 \int_0^\infty dm^2 E^C_{m^2}
\eeq

According to eq. \eqref{casimir_inequality}
\beq
I<E-E^C_0<E^C_{l^{-2}}-E^C_{\alpha/al}+I.
\eeq
$E^C_{l^{-2}}$ and $E^C_{\alpha/al}$ are both exponentially suppressed for small values of $l$. As a result, to leading order
\beq
E-E^C_0 = I = -l^2 \int_0^\infty dm^2 E^C_{m^2}.
\eeq
Note that the above equation is valid for any spacetime dimension. This shows that in any dimension, the modification to Casimir force is at order $l^2$ and has an opposite sign. Calculating the right hand side for $D=2$ (Appendix \ref{app1}), we get
\beq
E = -\frac{\pi}{24 a }+\frac{\pi^3 l^2 }{360 a^3}.
\eeq

\subsection{Numerical evaluation}\label{numerical_evaluation}

Here, we evaluate the Casimir energy numerically for a {\it finite} value of $N$. Instead of directly calculating the Casimir energy from eq. \eqref{casimir_disc}, we will use a different method that makes it easier to approach large values of $N$. 

As we have argued before, for any value of $N_0$,
\beq\label{casimir_numerics}
\sum_{k=1}^{N_0}E^C_{k \Delta}-E^C_{u_k \Delta}<E-E_0^C<E^C_{N\Delta}-E^C_{N_0\Delta}+\sum_{k=1}^{N_0}E^C_{k \Delta}-E^C_{u_k \Delta}.
\eeq
We will calculate the sum $\sum_{k=1}^{N_0} E^C_{k \Delta}-E^C_{u_k \Delta}$ and use it as an approximation to $E-E_0^C$. In order to do this, we have to choose the value of $N_0$ to be big enough such that $E^C_{N_0\Delta}$ becomes negligible, at the same time small enough such that the sum can be handled relatively easy. For this purpose, we choose

\begin{figure}[t]
\includegraphics[width=\textwidth]{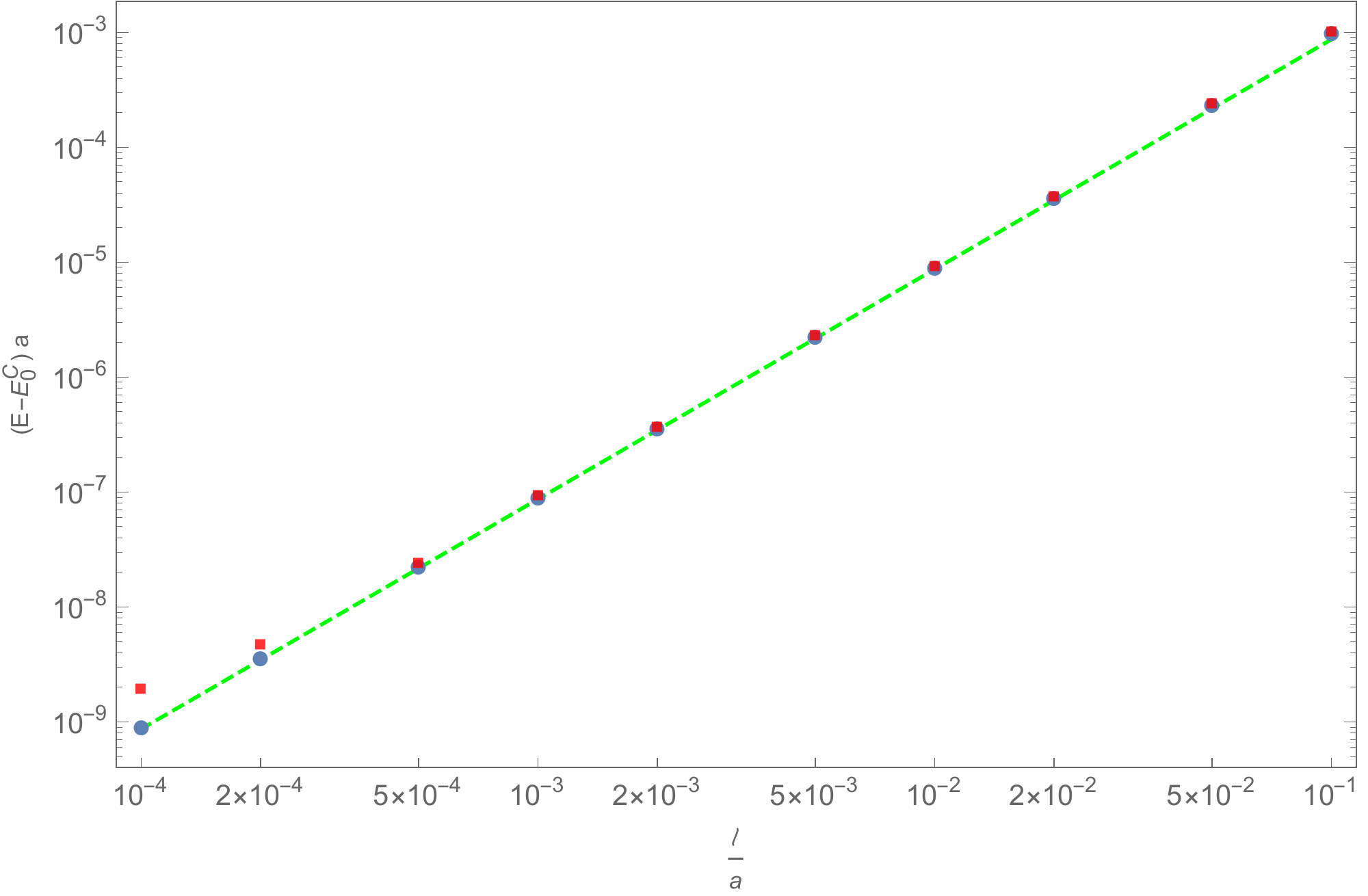}
\caption{For parameters specified in Sec. \ref{numerical_evaluation}, red squares (blue circles) represent the right hand side (left hand side) of eq. \eqref{casimir_numerics} multiplied by $a$, for different values of $l/a$. The real value of $E-E_0^C$ lies between them according to eq. \eqref{casimir_numerics}. The green dashed line shows the analytical value $\left(\frac{\pi^3l^2}{360a^2}\right)$. Note that circles and squares are on top of each other and on the green line for $l/a\gtrsim 5\times 10^{-4}$. For smaller values of $l/a$, we need to choose a bigger value for $ N_0$ to make the separation between the left hand and right hand side of eq. \eqref{casimir_numerics} smaller.}\label{fig:casimir_numerics}
\end{figure}

\beq
N_0\Delta=100/a^2\rightarrow N_0=100 N l^2/a^2.
\eeq 
Since the significant contribution to the sum comes from masses between $0$ and $1/a$, we fix $\Delta=\frac{1}{1000a^2}$ to sample this range with 1,000 points. As a result we choose the following parameters
\bea
&&N=\frac{1}{\Delta l^2}=1000 \frac{a^2}{l^2}\\
&&N_0=10^5.
\eea

Figure \ref{fig:casimir_numerics} shows the result of the numerical evaluation. In this figure we have presented the left hand side and right hand side of eq. \eqref{casimir_numerics}. The true value of the modification to Casimir energy lies between these two values.  Figure \ref{fig:casimir_numerics} shows that the result is consistent with the analytical derivations.

\section{Concluding remarks}

In this paper, we have covered different methods to calculate the modification to Casimir force between two parallel planes for a nonlocal field theory. In particular, we have shown that the modification is suppressed by $l^2/a^2$ in any dimension and has an opposite sign. In other words, the Casimir force is reduced due to nonlocality effects. 

One may compare this result with the Casimir force calculation presented in \cite{Saravani:2018rwm}, where the leading order modification is appeared at order $l$ and not $l^2$. As we have discussed in the introduction, the case in \cite{Saravani:2018rwm} is a different physical setup. There, the scalar field is assumed to vanish only at boundaries, $\phi(\x=0)=\phi(\x=a)=0$. Here, we have imposed a more stringent boundary conditions, $\phi(\x<0)=\phi(\x>a)=0$. For a local scalar field, both setups lead to the same value for Casimir force.
However, a nonlocal field theory is sensitive to the distinction between these physically different cases. From experimental point of view, it is safe to assume that the thickness of the ``walls'' is much larger than the nonlocality length scale. Thus, the result presented here seems more relevant for experiments.

\pagebreak

\bibliographystyle{jhep}
\bibliography{Casimir}

\pagebreak

\appendix
\section{Modification to Casimir energy}\label{app1}

In this section, we present a rather simple way to evaluate 
\beq
I = -l^2 \int_0^\infty dm^2 E^C_{m^2},
\eeq
in 1+1 dimensions,
where we have
\beq
E^C_{m^2} = \frac{1}{2\pi}\int_0^\infty dk~ \ln\left[1-e^{-2\sqrt{k^2+m^2}a}\right].
\eeq
Then,
\bea
I &=& -\frac{l^2}{2\pi} \int_0^\infty dm^2 dk ~ \ln\left[1-e^{-2\sqrt{k^2+m^2}a}\right]\notag\\
&=& -\frac{l^2}{2\pi a^3} \int_0^\infty dm^2 dk ~ \ln\left[1-e^{-2\sqrt{k^2+m^2}}\right]\notag\\
&=& \frac{l^2}{2\pi a^3} \int_0^\infty dk dm^2 ~ \sum_{n=1}^{\infty} \frac{e^{-2n\sqrt{k^2+m^2}}}{n}\label{log_expansion}\\
&=& \sum_{n=1}^{\infty} \frac{l^2}{2\pi a^3 n} \int_0^\infty dk dm^2 ~  e^{-2n\sqrt{k^2+m^2}}\notag\\
&=& \sum_{n=1}^{\infty} \frac{l^2}{2\pi a^3 n} \int_0^\infty dk ~ \frac{e^{-2nk}(1+2nk)}{2n^2}\label{m_integral}\\
&=& \sum_{n=1}^{\infty} \frac{l^2}{4\pi a^3 n^4}\label{k_integral}\\
&=& \frac{\pi^3l^2}{360a^3}.\label{n_sum}
\eea
where in eqs. \eqref{log_expansion}, \eqref{m_integral}, \eqref{k_integral} and \eqref{n_sum} we have performed the expansion of $\ln$, integral over $m^2$, integral over $k$ and sum over $n$, respectively.

\end{document}